\documentclass[aps,prl,reprint,groupedaddress,showpacs]{revtex4-1}

\usepackage{graphicx}
\usepackage{color} 

\begin{document}


\title{Role of transverse displacements in the formation of subaqueous barchan dunes\\
\textnormal{Accepted manuscript for Physical Review Letters, 121, 164503 (2018), DOI: 10.1103/PhysRevLett.121.164503}}



\author{Carlos A. Alvarez}
 \email{calvarez@fem.unicamp.br}
\author{Erick M. Franklin}
 \email{franklin@fem.unicamp.br}
 \thanks{Corresponding author}
\affiliation{%
School of Mechanical Engineering, UNICAMP - University of Campinas,\\
Rua Mendeleyev, 200, Campinas, SP, Brazil\\
}%


\date{\today}

\begin{abstract}
Crescentic shape dunes, known as barchan dunes, are formed by the action of a fluid flow on a granular bed. These bedforms are common in many environments, existing under water or in air, and being formed from grains organized in different initial arrangements. Although they are frequently found in nature and industry, details about their development are still to be understood. In a recent paper [C. A. Alvarez and E. M. Franklin, Phys. Rev. E 96, 062906 (2017)], we proposed a timescale for the development and equilibrium of single barchans based on the growth of their horns. In the present Letter, we report measurements of the growth of horns at the grain scale. In our experiments, conical heaps were placed in a closed conduit and individual grains were tracked as each heap, under the action of a water flow, evolved to a barchan dune. We identified the trajectories of the grains that migrated to the growing horns, and found that most of them came from upstream regions on the periphery of the initial heap, with an average displacement of the order of the heap size. In addition, we show that individual grains had transverse displacements by rolling and sliding that are not negligible, with many of them going around the heap. The mechanism of horns formation revealed by our experiments contrasts with the general picture that barchan horns form from the advance of the lateral dune flanks due to the scaling of migration velocity with the inverse of dune size. Our results change the way in which the growth of subaqueous barchan dunes is explained.
\end{abstract}


\maketitle


Dunes are the result of the interaction between granular matter and fluid flow. Crescentic shape dunes with horns pointing downstream, known as barchan dunes, are formed under one-directional flow and limited amount of available grains \cite{Bagnold_1, Herrmann_Sauermann, Hersen_3, Elbelrhiti}. When those conditions are present, barchan dunes are strong attractors, growing in different environments such as, for example, rivers, water ducts, Earth deserts, and on the surface of Mars \cite{Claudin_Andreotti, Parteli2}. For this reason, many studies were devoted to bed instabilities giving rise to barchans \cite{Kroy_C, Guignier, Parteli3, Khosronejad}, and to their equilibrium and minimum sizes \cite{Sauermann_1, Hersen_1, Andreotti_1, Hersen_2, Hersen_3, Kroy_B, Parteli, Andreotti_4, Franklin_8, Pahtz_1, Kidanemariam}.

Fewer studies were devoted to the growth and stability of horns, although they are one of the main features of a barchan dune. Khosronejad and Sotiropoulos \cite{Khosronejad} investigated numerically the instabilities on the surface of horns of grown barchans. They showed that transverse waves may appear and propagate over horns, giving rise to new barchans. In addition, they showed that the transverse waves and new barchans have related amplitude and wavelength. With their minimal models, Hersen \cite{Hersen_3} and Schw{\"a}mmle and Herrmann \cite{Schwammle} investigated numerically the formation of aeolian barchans from different initial shapes. In both their models, lateral diffusion was included. Hersen \cite{Hersen_3} argued that the diffusive effect was due to the lateral displacement of reptons: when salting grains impact onto the bed they cause the transverse displacement of grains by reptation. He proposed that aeolian barchans can be modeled as longitudinal 2D slices with lateral sand flux among them due mainly to reptation (diffusion), but also to air entrainment and slope effects; therefore, because the slice celerity varies with the inverse of its size \cite{Bagnold_1, Kroy_C, Andreotti_1}, horns grow mainly with grains originally in the lateral flanks of the initial heap, the other grains coming by lateral displacements from central regions of the pile. Although this picture is generally accepted for aeolian dunes, it has never been experimentally verified. Experiments shedding light on trajectories of grains over barchan dunes, including transverse displacements, are important to better understand the formation of barchans and to develop new continuous models (that are still necessary given the large number of grains involved in the problem). In a recent paper \cite{Alvarez}, we studied the formation of single barchans from initially conical heaps by investigating the growth of horns. For the length of horns as a function of time, we showed the existence of an initially positive slope, corresponding to its development, and a final plateau, corresponding to an equilibrium length for horns. We proposed the characteristic times $0.5 t_c$ for the growth and $2.5 t_c$ for equilibrium of barchans, where $t_c$ is a characteristic time for the displacement of barchans computed as the length of the bedform divided by its celerity $C$, and described in Ref. \cite{Alvarez}.

In this Letter, we present an experimental investigation on the movement of grains during the formation of horns in subaqueous dunes. Measurements at the grain scale over the entire bedform, essential to understand the dynamics of dunes, are presented here for the first time in the formation of barchan dunes. We show two new results: (i) grains forming the horns come mostly from upstream regions on the periphery of the initial heap; and (ii) transverse displacements by rolling and sliding are important for the growth of horns in the subaqueous case. In this way, the general picture for aeolian dunes that presumes that horns grow mainly with grains originally in the lateral flanks of the initial heap do not apply for subaqueous barchans. While ballistic saltons and reptons exist over aeolian dunes \cite{Andreotti_5}, they do not exist in the present subaqueous case, where grains move by rolling and sliding over each other (see Supplemental Material for a movie showing the development of a barchan dune from an initially conical heap \cite{Supplemental}).

The experimental device used is the same as that in Ref. \cite{Alvarez}, consisting of a water reservoir, centrifugal pumps, a flow straightener, a 5-$m$-long closed-conduit channel, a settling tank, and a return line. The channel test section was 1 $m$ long, started 40 hydraulic diameters downstream of the channel inlet and had a rectangular cross section (width = 160 $mm$ and height 2$\delta$ = 50 $mm$). Controlled grains were poured in the test section, which was previously filled with water, forming conical heaps that were afterward deformed into a barchan shape by the imposed water flow. The displacements of grains were filmed with a high-speed camera placed above the channel. The layout of the experimental device, a photograph of the test section, and an image of scanning electron microscopy of the used grains are shown in the Supplemental Material \cite{Supplemental}. Fig. \ref{fig:barchan_formation} shows top views of an initially conical heap deformed into a crescentic shape by the water flow, where $R$ is the radius of the initial pile. It was defined as the maximum radius with origin at the centroid and that do not contain void regions. The experimental conditions and times are described in the figure caption.

\begin{figure}[b]
\includegraphics[width=\linewidth]{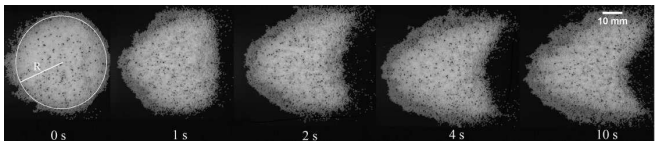}
\caption{Top views of an initially conical heap deformed by the water flow at different times (shown below each frame). The water flow is from left to right, $R$ is the radius of the initial pile, and black spots are the tracers. $Re$ = $1.82 \cdot 10^4 $ and the heap initial mass was 6.2 g.}
	\label{fig:barchan_formation}
\end{figure}

The tests were performed with tap water at temperatures within 24 and 26 $^{\circ}$C and round glass beads ($\rho_s$ = 2500 kg/m$^3$) with $0.40$ mm $\leq\,d\,\leq$ $0.60$ mm, where $\rho_s$ and $d$ are, respectively, the density and diameter of glass beads. In order to facilitate the tracking of moving grains, 2$\%$ of them were tracers, i.e., glass beads of different color but same diameter and surface characteristics as the other grains. The tracers were not painted, they were made of colored glass. The cross-section mean velocities $U$ were 0.243, 0.294 and 0.364 m/s, corresponding to Reynolds numbers based on the channel height $Re=\rho U 2\delta /\mu$ of $1.21 \cdot 10^4$, $1.47 \cdot 10^4 $ and $1.82 \cdot 10^4 $, respectively, where $\mu$ is the dynamic viscosity and $\rho$ the density of the fluid. The shear velocities on the channel walls were computed from velocity profiles acquired by a two-dimensional particle image velocimetry (2D-PIV) device and were found to follow the Blasius correlation \cite{Schlichting_1}. They correspond to 0.0149, 0.0177 and 0.0213 m/s for the three flow rates employed. The initial heaps were formed with 6.2 and 10.3 g of glass beads, corresponding to initial volumes of 4.1 and 6.9 cm$^3$, respectively, and to $R$ of 2.6 and 3.2 cm, respectively. For each experimental condition we performed a minimum of 3 test runs.

\begin{figure}[b]
\includegraphics[width=0.8\linewidth]{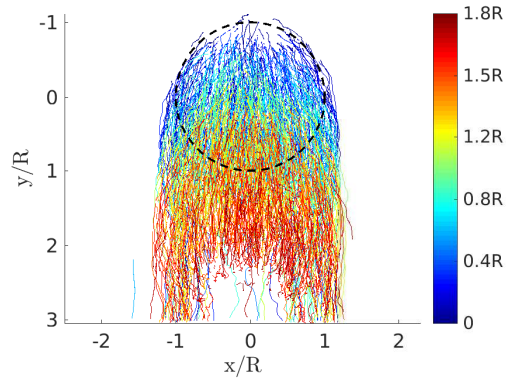}
\caption{Trajectories of all marked grains during the growth of a barchan dune. The water flow is from top to bottom. $Re$ = $1.82 \cdot 10^4 $ and the heap initial mass was equal to 6.2 g. The black dashed circle represents the initial pile of radius R.}
	\label{fig:all_tracers}
\end{figure}

With an image processing code based on Refs. \cite{Kelley} and \cite{Alvarez2}, we identified the centroids of tracer grains and tracked them along the movie frames. Fig. \ref{fig:all_tracers} shows the trajectories of all marked grains during the growth of a barchan dune from the initial conical pile. Because the pile moves while the horns grow, the color of pathlines in Fig. \ref{fig:all_tracers} changes from blue to red to indicate different positions of the pile centroid: they are blue at the initial position and red at the final position. The scaling bar in the figure shows the respective values of $r_c - r_0$ in terms of $R$, where $r_c$ and $r_0$ are, respectively, the instantaneous and initial positions of the pile centroid, and the black dashed circle represents the initial pile. We note from this figure that grains have significant transverse movements, and that part of the grains going to horns comes from the upstream periphery of the pile, describing circular paths.

\begin{figure}[b]
\includegraphics[width=0.9\linewidth]{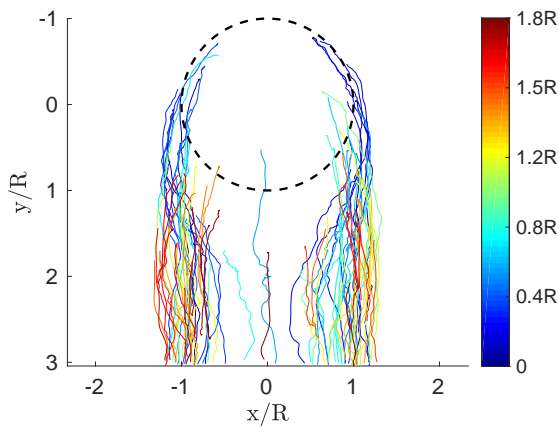}
\caption{Trajectories of marked grains that migrated to horns during the growth of a barchan dune. The water flow is from top to bottom. $Re$ = $1.82 \cdot 10^4 $ and the heap initial mass was 6.2 g. The black dashed circle represents the initial pile of radius R.}
	\label{fig:horn_tracers}
\end{figure}

Fig. \ref{fig:horn_tracers} shows the trajectories of marked grains that migrated to horns during the growth of a barchan dune (see Supplemental Material \cite{Supplemental} for trajectories of other cases). As in Fig. \ref{fig:all_tracers}, pathlines are blue at the pile initial position and red at the barchan final position. From Fig. \ref{fig:horn_tracers}, it is noticeable that a significant part of grains going to horns was originally in the upstream region of the pile periphery, and that their path is described approximately by an arc of circumference.

\begin{figure}[b]
\includegraphics[width=0.9\linewidth]{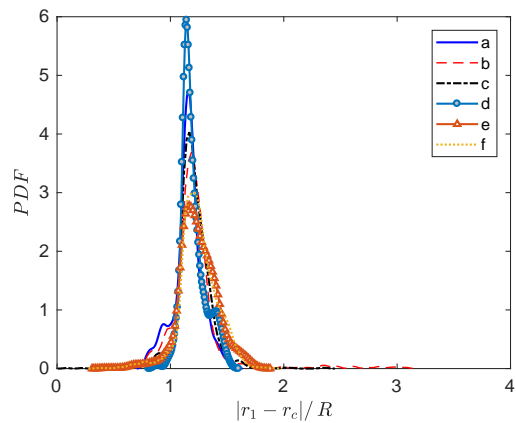}
\caption{PDF of the original position of the grains that migrated to horns during the growth of a barchan dune. A kernel smoothing function was used to plot the PDF \cite{Bowman_Azzalini}. The cases a, b and c correspond to $Re$ = $1.21 \cdot 10^4$, $1.47 \cdot 10^4 $ and $1.82 \cdot 10^4 $, respectively, and initial mass equal to 6.2 g. The cases d, e, and f correspond to $Re$ = $1.21 \cdot 10^4$, $1.47 \cdot 10^4 $ and $1.82 \cdot 10^4 $, respectively, and initial mass equal to 10.3 g.}
	\label{fig:pdf_origin_horns}
\end{figure}

\begin{figure}
   \begin{minipage}[c]{\columnwidth}
    \begin{center}
     \includegraphics[width=.75\linewidth]{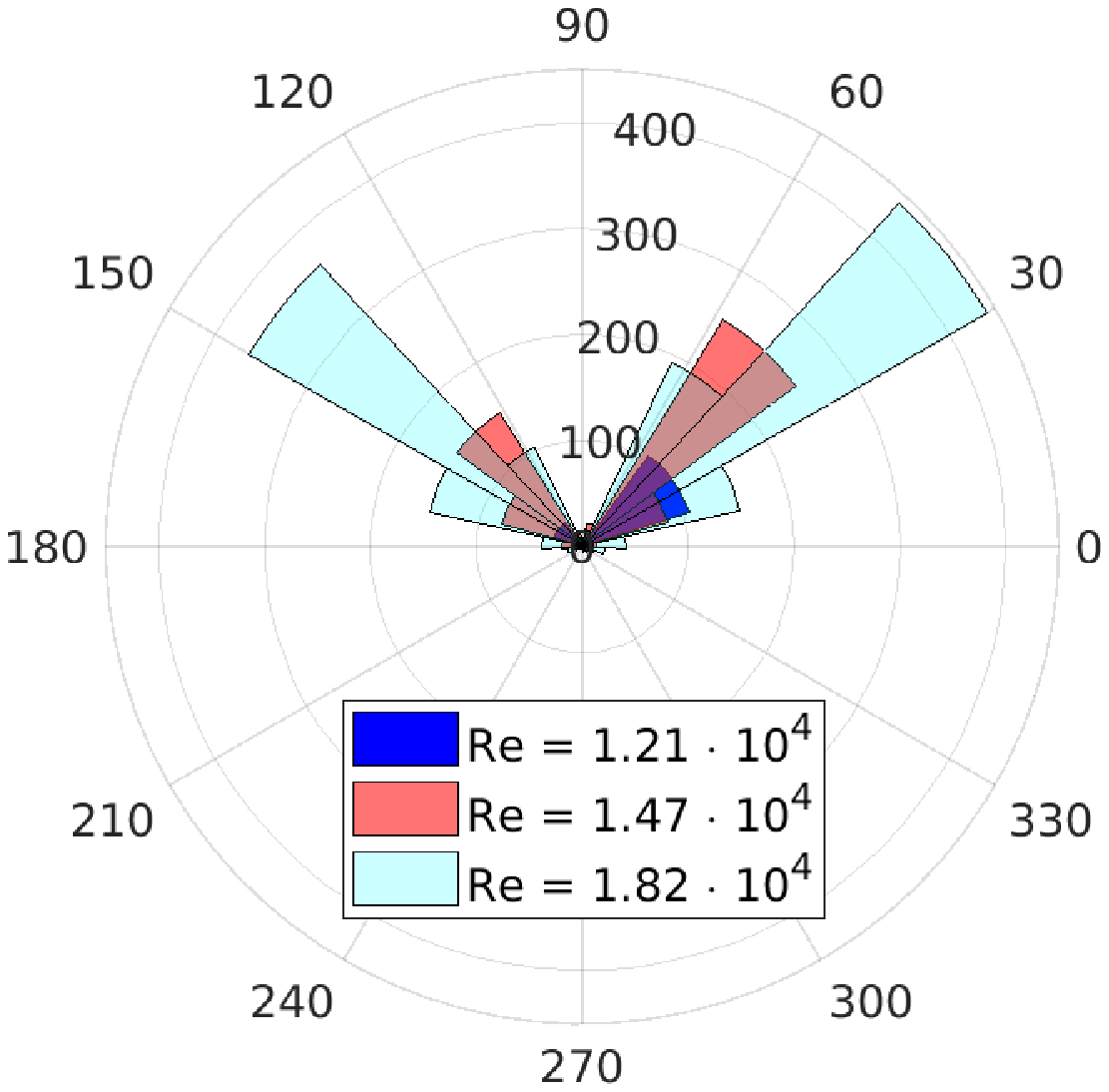}\\
		(a)
    \end{center}
   \end{minipage} \hfill
   \begin{minipage}[c]{\columnwidth}
    \begin{center}
      \includegraphics[width=.75\linewidth]{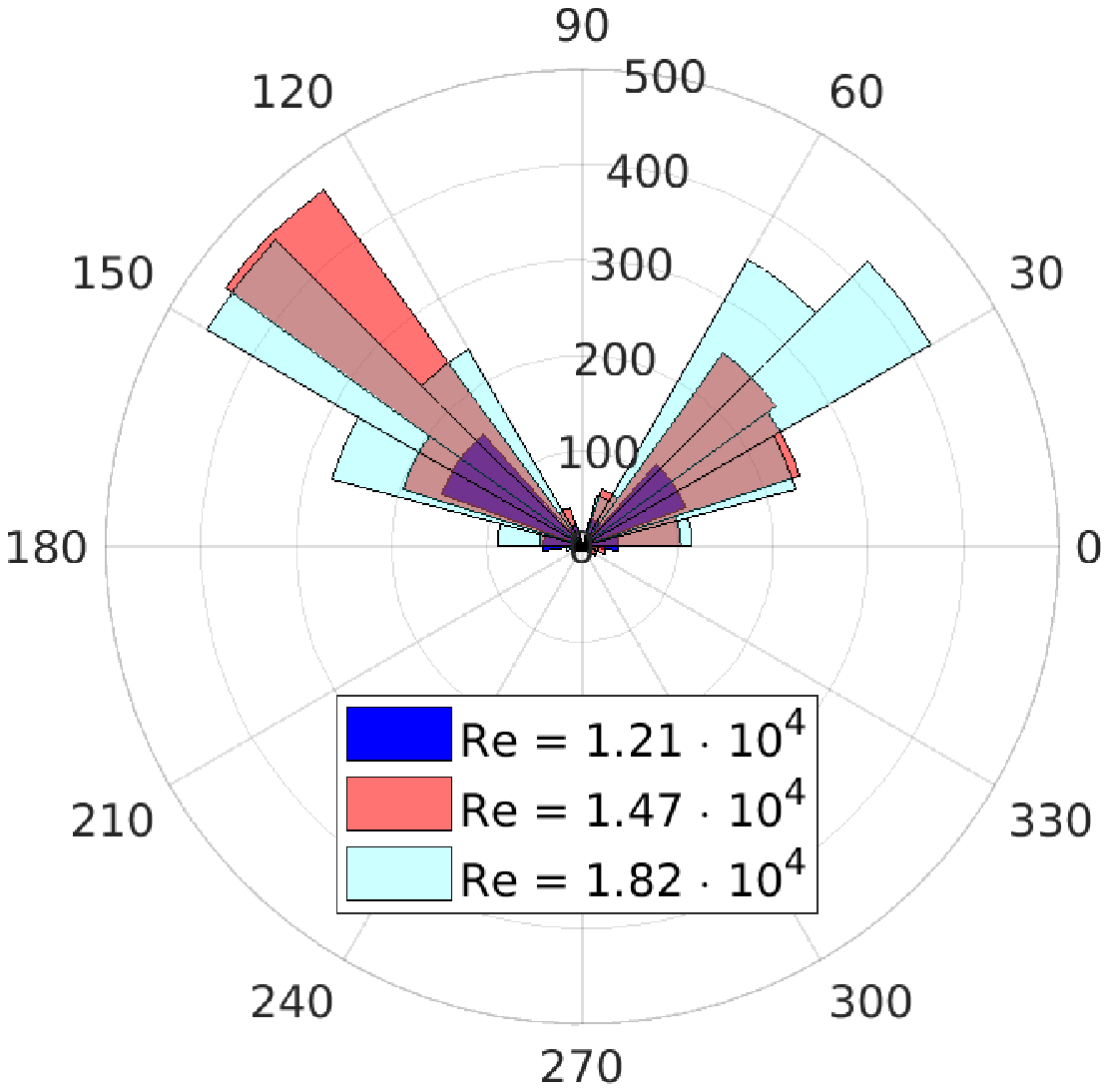}\\
		(b)
    \end{center}
   \end{minipage}
\caption{Frequency of occurrence of the initial position of grains migrating to the horns as a function of the angle with respect to the transverse direction (water flow direction is 270$^{\circ}$). (a) Initial mass equal to 6.2 g and (b) initial mass equal to 10.3 g. The tips of appearing horns point to angles of approximately 240$^{\circ}$ and 300$^{\circ}$.}
	\label{fig:origin_horns}
\end{figure}

In order to identify the exact upstream regions from where the grains migrate to the horns, we investigated the origin of grains based on the radial and angular positions. Fig. \ref{fig:pdf_origin_horns} presents the probability density function (PDF) of the initial position $r_1$ of grains migrating to the horns as a function of the normalized radial position $|r_1 - r_c|/R$, and Fig  \ref{fig:origin_horns} shows the frequency of occurrence of the initial position of the same grains as a function of the angle with respect to the transverse direction. We note from Figs. \ref{fig:pdf_origin_horns} and \ref{fig:origin_horns} that most part of the grains going to horns was originally on the periphery of the initial conical pile, with $|r_1 - r_c|/R$ $>$ 1 and angles between 15$^{\circ}$ and 60$^{\circ}$ and 120$^{\circ}$ and 165$^{\circ}$ with respect to the transverse direction. The asymmetries in Fig. \ref{fig:origin_horns} reflect the experimental dispersions of our experiments. Those grains are entrained directly by the water flow, moving without reptation (see Supplemental Material for a movie showing the movement of grains \cite{Supplemental}). The circular path described by the grains going to horns (Fig. \ref{fig:horn_tracers}) is completely different from the paths observed in the aeolian case, where salting grains effectuate ballistic flights in the wind direction, impacting in many instances onto the dune surface, and reptation exists \cite{Hersen_3, Schwammle, Andreotti_5}.

Finally, we measured the total distance traveled by each grain that migrated to the horns and computed the mean traveled distance $L_{mean}$. We found that

\begin{equation}
22 < L_{mean}/L_{drag} < 30
\end{equation}

\noindent for the 6.2 g piles and

\begin{equation}
18 < L_{mean}/L_{drag} < 22
\end{equation}
 
\noindent for the 10.3 g piles, where $L_{drag}$ = $( \rho_s / \rho ) d$ is an inertial length proportional to the length for the stabilization of sand flux \cite{Hersen_1}. The diameter of the initial conical pile was approximately 42$L_{drag}$ and 49$L_{drag}$ for the 6.2 and 10.3 g piles, respectively; therefore, a large part of grains migrating to horns travel distances of the order of the pile radius until reaching the horns. In the aeolian case, $L_{mean}$ is unknown, and future work can help to include some of the present findings in aeolian models.

Lajeunesse et al. \cite{Lajeunesse}, Seizilles et al. \cite{Seizilles} and Penteado and Franklin \cite{Penteado} investigated the displacements of individual grains on plane granular beds under laminar \cite{Seizilles} and turbulent \cite{Lajeunesse, Penteado} water flows. They found that the transverse velocity of particles is distributed around a zero mean value, and that the streamwise velocity is a function of the excess of shear stress with respect to a threshold value \cite{Lajeunesse, Penteado}. Seizilles et al. \cite{Seizilles} investigated in particular the transverse displacements of bed load particles, and proposed a Fickian diffusion mechanism across the flow direction. They found a diffusion length $\ell_d$ of approximately 0.030$d$, which gives $\ell_d/L_{drag} \, \approx$ 0.012, three orders of magnitude smaller than the values found for $L_{mean}$ in the present study. This corroborates our argument about the absence of diffusion in the trajectories of grains migrating to the horns. Ref. \cite{Penteado}, that used glass beads within the range of diameters of the present study, measured 3 $<$ $L_{mean}/L_{drag}$ $<$ 9. Values in the present study are 3 times greater than those of Ref. \cite{Penteado} probably due to the water acceleration around the granular pile and to the fact that some grains move directly over the bottom wall of the channel.

In conclusion, the trajectories of grains going to the horns during the growth of a subaqueous barchan have significant transverse components, with a great part of grains being originally in upstream regions of the pile periphery. The grains move by rolling and sliding, being entrained directly by the water flow and traveling distances of the order of the pile size. This is different from the aeolian case, where saltons effectuate ballistic flights in the wind direction and the transverse displacements are due in part to reptation. However, the relative importance of reptation to transverse displacements in the aeolian case is still to be determined. The present results change our understanding on the formation of subaqueous barchans.

Carlos A. Alvarez is grateful to SENESCYT (grant no. 2013-AR2Q2850) and to CNPq (grant no. 140773/2016-9). Erick M. Franklin is grateful to FAPESP (grant no. 2016/13474-9), to CNPq (grant no. 400284/2016-2) and to FAEPEX/UNICAMP (grant no. 2210/18) for the financial support provided.

\bibliography{references}

\end{document}